\documentclass[conference]{IEEEtran}

\IEEEoverridecommandlockouts

\usepackage{cite}
\usepackage{amsmath,amssymb}
\usepackage{graphicx}
\usepackage{booktabs}
\usepackage{multirow}
\usepackage{url}
\usepackage{xcolor}
\usepackage{microtype}
\usepackage[colorlinks=true,linkcolor=blue,citecolor=blue,urlcolor=blue]{hyperref}

\newcommand{\our}{\textsc{HonestAffinity}}

\title{HonestAffinity: Leak-Aware Evaluation of Protein and Pocket Priors for Binding Affinity Prediction}

\author{%
  \IEEEauthorblockN{%
    Junhao Wei\textsuperscript{1},
    Baili Lu\textsuperscript{1,2},
    Zhenhong Peng\textsuperscript{2},
    Wanyan Li\textsuperscript{2},
    Zhirong Huang\textsuperscript{3},
    Yanxiao Li\textsuperscript{1},
    Yifu Zhao\textsuperscript{1},\\
    Dexing Yao\textsuperscript{1},
    Haochen Li\textsuperscript{1},
    Xudong Ye\textsuperscript{1},
    Sio-Kei Im\textsuperscript{4},
    Yapeng Wang\textsuperscript{1},
    Xu Yang\textsuperscript{1,*}
  }
  \IEEEauthorblockA{%
\textsuperscript{1}\textit{Faculty of Applied Sciences, Macao Polytechnic University, Macao SAR, China}\\
    \textsuperscript{2}\textit{College of Animal Science and Technology, Zhongkai University of Agriculture and Engineering, Guangzhou, China}\\
    \textsuperscript{3}\textit{College of Animal Science, South China Agricultural University, Guangzhou, China}\\
    \textsuperscript{4}\textit{Macao Polytechnic University, Macao SAR, China}\\
    \textsuperscript{*}Corresponding author: Xu Yang \quad \texttt{xuyang@mpu.edu.mo}%
  }
}

\begin{document}
\maketitle

\begin{abstract}
Sequence-based deep learning offers a scalable alternative to structure-based scoring for protein--ligand binding affinity prediction, but its progress is difficult to interpret when architectural priors are evaluated only on canonical PDBbind-style splits that leak protein--ligand similarity classes across folds. We present \our{}, a compact 1D-input affinity predictor used to isolate two common priors under a leak-aware protocol: frozen ESM-2 (650M) per-residue protein embeddings and a learned binary pocket-position marker. The same multi-scale convolutional / single-layer Transformer template is evaluated in three variants: \our{}-\textsc{Pocket} (ESM-2 + pocket marker), \our{}-\textsc{NoPocket} (ESM-2 only), and \our{}-\textsc{Pocket-NoESM} (21-vocabulary residue embedding + pocket marker). All three train on 11{,}513 LP-PDBBind complexes in $\sim$3 GPU-hours on a single Tesla V100. We benchmark them against five reproduced baselines on the LP-PDBBind 3-tier no-leak hold-out, CASF-2016, and a 161-complex CASF-2016 non-train subset, reporting every method as mean$\pm$std over three random seeds. The central finding is not a uniformly best prior, but a \emph{split-conditioned reversal}: \our{}-\textsc{Pocket} achieves the best mean Pearson R on validation, CASF-2016, and CASF-2016 non-train, whereas \our{}-\textsc{Pocket-NoESM} achieves the best mean Pearson R on every strict LP no-leak tier (\texttt{test\_cl1}--\texttt{cl3}), including relative to our ESM-equipped variants. Thus, both the pocket marker and the 1280-dimensional ESM-2 input help on familiar/CASF-style splits but reduce Pearson R on strict no-leak tiers. We argue that binding-affinity models should report paired canonical and leak-proof ablations, and that deployment-regime-matched variants describe such reversals more faithfully than a single global default. An anonymous code snapshot and preprocessing scripts are linked in the first-page footnote; pretrained checkpoints will be released upon acceptance.
\end{abstract}

\begin{IEEEkeywords}
Protein--ligand affinity, drug discovery, protein language model, ESM-2, leak-proof evaluation, LP-PDBBind, CASF-2016.
\end{IEEEkeywords}

\section{Introduction}
\label{sec:intro}

Estimating the binding affinity of a small-molecule ligand to a target protein is a central task in early-stage AI-driven drug discovery (AIDD): a fast, accurate scoring function turns a virtual library of millions of compounds into a tractable ranked shortlist. Two families dominate the literature. \emph{Structure-based} scorers consume 3D complex geometry (e.g., voxel grids, atomic graphs, equivariant networks) and are accurate when crystal poses or high-quality docks are available. \emph{Sequence-based} scorers consume only the protein amino-acid sequence and the ligand SMILES, sidestepping structure availability and enabling library-scale screening for targets without solved structures.

Recent sequence-based models have narrowed the apparent gap to structure-based scoring, but the reported progress remains difficult to interpret. One source of uncertainty is the protein representation itself. Models such as DeepDTA~\cite{deepdta}, DeepDTAF~\cite{deepdtaf}, and recent multi-scale matrix-product architectures~\cite{dba2025} typically encode amino acids with a learned 21-token embedding, whereas billion-parameter protein language models such as ESM-2~\cite{esm2,esm1b} already encode evolutionary co-conservation, secondary-structure cues, and binding-site signatures. Whether this pretrained protein prior improves binding-affinity prediction under strict no-leak evaluation remains unclear.

A second source of uncertainty is the evaluation split. The default PDBbind general/refined/core protocol shares protein--ligand similarity classes between train and test, so models that overfit to those classes can still report strong CASF numbers while failing on dissimilar targets. The leak-proof LP-PDBBind~\cite{lppdbbind} reorganization addresses this issue and exposes 30--50\% drops for many state-of-the-art scorers, but few sequence-only papers have re-evaluated their models under this regime.

This paper addresses both issues by using \our{} as a controlled testbed for protein and pocket priors. \our{} is a 1D-input affinity predictor in which the protein branch can consume pre-cached ESM-2 (650M) per-residue embeddings and/or an optional learned binary pocket-position marker. We study \emph{three} variants of the same architectural template: \our{}-\textsc{Pocket} (ESM-2 plus pocket marker; familiar targets where pocket annotation is reliable), \our{}-\textsc{NoPocket} (ESM-2 only, no pocket annotation; targets with a sequence and a SMILES), and \our{}-\textsc{Pocket-NoESM} (21-vocabulary protein label embedding plus pocket marker; highest mean R in our strict LP no-leak tiers). The rest of the network is deliberately compact: a multi-scale 1D-convolution encoder, a residual block, a single Transformer layer, a mirrored SMILES branch, and a matrix-product compatibility score followed by an MLP head. All three variants train on a single GPU in $\sim$3 hours.

We benchmark all three \our{} variants on the LP-PDBBind 3-tier hold-out, on CASF-2016, and on the 161 CASF-2016 complexes that are not assigned to the LP-PDBBind training split, against five competitive baselines spanning the 1D-input (DeepDTA), voxel-based (Pafnucy~\cite{pafnucy}), pocket-aware sequence (DeepDTAF), attention-based (DEAttentionDTA~\cite{deattentiondta}), and multi-scale matrix-product~\cite{dba2025} families. All numbers in this paper are reported as mean$\pm$std across three random seeds for every \our{} variant and every reproduced baseline. Our contributions are:

\begin{itemize}
\item \textbf{A leak-aware evaluation of protein and pocket priors for affinity prediction} on the LP-PDBBind 3-tier hold-out plus CASF-2016 and a CASF-2016 non-train subset (161 entries not in LP train), with three-seed mean$\pm$std reporting for every method.
\item \textbf{A split-conditioned reversal that recurs across two unrelated components.} Two structurally unrelated input choices---a binary pocket-position marker and the 1280-dimensional frozen ESM-2 embedding---both \emph{help} the canonical / familiar splits (val, CASF-2016, CASF-2016 non-train) and \emph{reduce performance} on every strict LP no-leak tier. Single-regime reporting would conceal both reversals; we argue that the pattern is a property of the \emph{evaluation regime} rather than of either component, and motivates paired canonical / leak-proof ablations as a routine sanity check.
\item \textbf{Three deployment-regime-matched variants of \our{} as the practical consequence.} Rather than collapse the reversal to a single best model, we define three variants of the same architecture, each matched to a distinct deployment regime: \our{}-\textsc{Pocket} for familiar / annotated targets (highest mean Pearson R on val, CASF-2016, CASF-2016 non-train), \our{}-\textsc{Pocket-NoESM} for strict LP-style targets with pocket annotations (highest mean Pearson R on \texttt{test\_cl1}--\texttt{cl3}), and \our{}-\textsc{NoPocket} for targets without pocket annotations.
\end{itemize}

\section{Related Work}
\label{sec:related}

\subsection{Sequence- and voxel-based affinity models}
DeepDTA~\cite{deepdta} introduced parallel CNNs over protein and SMILES character sequences and remains a canonical sequence baseline. Pafnucy~\cite{pafnucy} discretizes the protein--ligand complex into a $21^3$ voxel grid with 19 atom-type channels and applies a 3D CNN. DeepDTAF~\cite{deepdtaf} adds an explicit pocket-sequence channel and concatenates protein/pocket/ligand features into a deeper convolutional pipeline. DEAttentionDTA~\cite{deattentiondta} introduces a dynamic-embedding self-attention layer that re-weights protein tokens by ligand context. CAPLA~\cite{capla} uses cross-attention over residue--atom features. Multi-scale convolution combined with matrix-product affinity scoring is taken up by recent work~\cite{dba2025}, one of the sequence-only baselines against which we compare.

\subsection{Protein language models in molecular tasks}
ESM-2~\cite{esm2} and its earlier variants~\cite{esm1b} have been shown to encode tertiary-structure cues recoverable by linear probes, and have been integrated into single-sequence structure prediction pipelines and into protein--protein and protein--ligand interface predictors. Comparable ligand-side foundation models include Uni-Mol~\cite{unimol} and MolFormer~\cite{molformer}. We inject frozen ESM-2 embeddings into the multi-scale convolutional / matrix-product scoring template that has been successful on CASF-style evaluation, then test whether the resulting protein prior still helps under the LP-PDBBind regime.

\subsection{Data-quality and leak-proof benchmarks}
The PDBbind database~\cite{pdbbind,casf2016} has long anchored protein--ligand affinity benchmarks, but the standard general/refined/core splits permit leakage of similar protein--ligand pairs across folds. The leak-proof LP-PDBBind~\cite{lppdbbind} reorganization clusters the database and exposes three increasingly strict no-leak test tiers; reported drops on this benchmark have been $0.10$--$0.20$ in Pearson R for several published scorers. We adopt LP-PDBBind as our primary evaluation, complemented by CASF-2016 for comparability with prior work.

\section{Method}
\label{sec:method}

\begin{figure*}[t]
  \centering
  \includegraphics[width=0.95\linewidth]{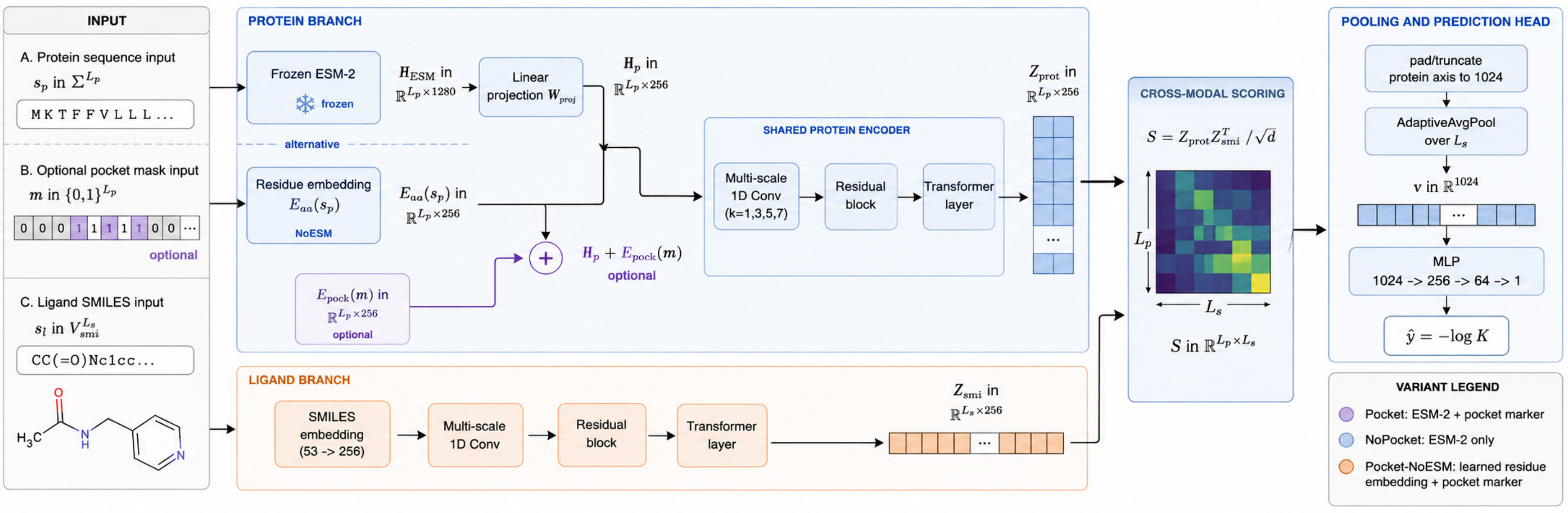}
  \caption{\textbf{Architecture of \our{}.} The schematic shows the shared 1D-input template and the optional paths that define the three deployment variants. The protein branch uses either frozen ESM-2 (650M) per-residue embeddings projected from 1280 to $d{=}256$, or, for \our{}-\textsc{Pocket-NoESM}, a learned residue embedding. When pocket annotations are available, a learned pocket-marker embedding is added to the protein representation; \our{}-\textsc{NoPocket} omits this marker. The ligand branch encodes tokenized SMILES with a learned 53-token embedding and the same multi-scale 1D-convolution / residual / Transformer pattern. The scoring head forms $\mathbf{S}=\mathbf{Z}^{\mathrm{prot}}{\mathbf{Z}^{\mathrm{smi}}}^{\!\!\top}/\sqrt{d}$; the prediction head pads/truncates the protein axis to 1024, averages over $L_s$, and applies an MLP ($1024\to256\to64\to1$) to predict $\hat{y}=-\log K$.}
  \label{fig:pipeline}
\end{figure*}

\subsection{Overview}
Fig.~\ref{fig:pipeline} summarizes \our{}. The \our{}-\textsc{Pocket} variant takes three inputs for a complex: the full-length protein amino-acid sequence $\mathbf{s}_p\in\Sigma^{L_p}$, a binary pocket-position mask $\mathbf{m}\in\{0,1\}^{L_p}$ derived from the deposited pocket residues, and the ligand SMILES $\mathbf{s}_\ell\in\mathcal{V}_{\mathrm{smi}}^{L_s}$. The \our{}-\textsc{NoPocket} variant omits $\mathbf{m}$ and consumes only $(\mathbf{s}_p, \mathbf{s}_\ell)$. The \our{}-\textsc{Pocket-NoESM} variant follows \our{}-\textsc{Pocket} but replaces the projected ESM-2 stream with a learned 21-vocabulary amino-acid embedding (defined formally in Section~\ref{sec:protenc}). All three variants produce a scalar affinity $\hat{y}=-\log K$.

\subsection{PLM-augmented protein encoder}
\label{sec:protenc}
Per-residue ESM-2 (650M) embeddings $\mathbf{H}_p\in\mathbb{R}^{L_p\times1280}$ are precomputed once and cached on disk (24\,GB for PDBbind v2020R1, $\sim$1.3\,MB per protein). A learned linear projection $\mathbf{W}_{\mathrm{proj}}\in\mathbb{R}^{1280\times d}$ with $d{=}256$ reduces the embedding dimension. For \our{}-\textsc{Pocket}, a pocket-position embedding $\mathbf{E}_{\mathrm{pock}}\in\mathbb{R}^{2\times d}$ is added \emph{additively} at each residue:
\begin{equation}
\tilde{\mathbf{H}}_p[i] = \mathbf{W}_{\mathrm{proj}}\mathbf{H}_p[i] + \mathbf{E}_{\mathrm{pock}}[m_i].
\end{equation}
For \our{}-\textsc{NoPocket} the second term is omitted ($\tilde{\mathbf{H}}_p[i] = \mathbf{W}_{\mathrm{proj}}\mathbf{H}_p[i]$); no pocket annotation is consumed at any stage of training or inference. For \our{}-\textsc{Pocket-NoESM} the projected ESM-2 stream is replaced with a learned 21-vocabulary amino-acid embedding $\mathbf{E}_{\mathrm{aa}}\in\mathbb{R}^{21\times d}$ ($\tilde{\mathbf{H}}_p[i] = \mathbf{E}_{\mathrm{aa}}[s_{p,i}] + \mathbf{E}_{\mathrm{pock}}[m_i]$), keeping the pocket marker; this matches the protein-token regime of pre-PLM 1D-input baselines while preserving every other architectural choice and our pocket-marker mechanism.
The augmented features are passed through (a) four parallel 1D convolutions with kernels $\{1,3,5,7\}$ and channel widths $\{32,32,64,128\}$ projected back to $d$, (b) a residual ResNet block (Linear$\to$ReLU$\to$LayerNorm with skip), and (c) a single Transformer encoder layer~\cite{transformer} with $h{=}8$ heads of $d_k{=}16$. Padding positions are masked in self-attention. The output is a contextual protein representation $\mathbf{Z}^{\mathrm{prot}}\in\mathbb{R}^{L_p\times d}$.

\subsection{SMILES sequence encoder}
The ligand branch mirrors the protein branch. SMILES tokens are mapped through a 53-token learned embedding to $d$ channels and processed by an identical multi-scale convolution / residual / Transformer stack, producing $\mathbf{Z}^{\mathrm{smi}}\in\mathbb{R}^{L_s\times d}$.

\subsection{Affinity scoring head}
The two streams are coupled via a matrix product:
\begin{equation}
\mathbf{S} \;=\; \frac{1}{\sqrt{d}}\;\mathbf{Z}^{\mathrm{prot}}\,{\mathbf{Z}^{\mathrm{smi}}}^{\!\!\top}\;\in\;\mathbb{R}^{L_p\times L_s},
\end{equation}
which produces a per-residue $\times$ per-token compatibility map. We zero-pad or truncate $\mathbf{S}$ to a fixed protein length of $1024$ and collapse the SMILES dimension with adaptive 1D average pooling, yielding a vector $\mathbf{v}\in\mathbb{R}^{1024}$. A 3-layer MLP ($1024\to256\to64\to1$) with PReLU activations and $0.5$ dropout on the two hidden layers maps $\mathbf{v}$ to the final scalar affinity prediction.

\subsection{Training and reporting protocol}
We train with mean-squared error on $-\log K$ targets using AdamW~\cite{adamw} (lr $10^{-4}$, weight decay $10^{-4}$, $\epsilon{=}10^{-6}$), a cosine annealing schedule~\cite{cosine} over 80 epochs (early-stopped with patience 20 on validation RMSE), batch size 32, and gradient clipping at norm 1.0. Training is performed in single precision (fp32) end-to-end. ESM-2 weights are frozen throughout. For reporting, every \our{} variant, every ablation, and every reproduced baseline is trained from scratch with three independent random seeds; we report the per-split mean$\pm$standard deviation across the three runs and use validation RMSE for model selection within each seed. No single best run is hand-picked.

All experiments run on a single NVIDIA Tesla V100-DGXS GPU (32\,GB HBM2) hosted on an Ubuntu 20.04 DGX Station (4 V100s available; only one card is used per training run); the host has 256\,GB system RAM. The codebase is implemented in PyTorch 2.1 with CUDA 11.8. ESM-2 (650M) embeddings are extracted once with the official \texttt{fair-esm} release and cached as compressed NumPy arrays on disk (24\,GB for PDBbind v2020R1). One full training run of \our{} takes approximately 3 hours on this hardware; inference is $\sim$10\,ms per complex once the embedding cache is warm.

\section{Datasets and Splits}
\label{sec:data}

\subsection{PDBbind v2020R1}
We use the v2020R1 reprocessed release (Aug.\ 2025), which contains 19{,}037 protein--ligand complexes. We extract amino-acid and pocket sequences from the supplied protein and pocket PDB files using BioPython~\cite{biopython}, with secondary structure assigned by DSSP~\cite{dssp} (used only for the CAPLA-style baseline; \our{} does not use SSE). Ligand SMILES are obtained via RDKit~\cite{rdkit} from the supplied SDF/MOL2 files. After preprocessing, 19{,}032 complexes have all four streams (protein, pocket, ligand, label) available; the remainder fail due to malformed coordinates or unparseable ligands. Affinity targets are $-\log K_d/K_i/\mathrm{IC}_{50}$ as recorded.

\subsection{LP-PDBBind 3-tier hold-out}
We adopt the leak-proof splits of~\cite{lppdbbind} on the v2020 entries:
\begin{itemize}
  \item \textbf{train}: 11{,}513 complexes
  \item \textbf{val}: 2{,}422 complexes
  \item \textbf{test\_cl1}: 4{,}286 complexes (loosest no-leak filter)
  \item \textbf{test\_cl2}: 2{,}179 complexes
  \item \textbf{test\_cl3}: 1{,}348 complexes (strictest filter)
\end{itemize}
The three test tiers progressively exclude PDB entries whose protein and/or ligand exceed similarity thresholds with the training set, providing a graded view of generalization.

\subsection{CASF-2016 and CASF-2016 non-train subset}
We additionally evaluate on CASF-2016 (285 complexes; the standard scoring-power benchmark~\cite{casf2016}), which contains some entries also present in LP-PDBBind \emph{train}. To separate CASF entries seen during LP training from those not used for training, we also report \emph{CASF-2016 non-train}: the 161 CASF-2016 entries that are \emph{not} assigned to LP-PDBBind \emph{train}. This subset is not a fully independent benchmark: in our split file, 42 of the 161 complexes overlap LP validation, 108 appear in at least one LP no-leak test tier, and 11 are outside the LP split table. We therefore use it only as a non-training CASF diagnostic. We do not include CASF-2013 because the canonical PDB list was not bundled in the v2020R1 distribution we used.

\section{Experiments}
\label{sec:exp}

\subsection{Baselines}
We re-implement and re-train five competitive baselines on identical LP-PDBBind splits:
\begin{itemize}
  \item \textbf{DeepDTA}~\cite{deepdta} --- parallel CNNs over protein and SMILES character sequences.
  \item \textbf{Pafnucy}~\cite{pafnucy} --- $21^3$ voxel grid with 19 atom-type channels, 3D CNN. We regenerated the 19-channel grids for all 19{,}032 PDBbind v2020R1 complexes.
  \item \textbf{DeepDTAF}~\cite{deepdtaf} --- protein, pocket and ligand convolutional towers with concatenation.
  \item \textbf{DEAttentionDTA}~\cite{deattentiondta} --- dynamic embedding plus self-attention re-weighting (Bioinformatics 2024).
  \item \textbf{Multi-scale matrix-product}~\cite{dba2025} --- a recent 1D-input multi-scale CNN--Transformer baseline reproduced from the published reference implementation.
\end{itemize}
All baselines use their reference hyper-parameters (no tuning to LP splits), are reproduced in single precision (fp32) on the same single NVIDIA Tesla V100 (32\,GB) GPU as \our{}, and share our preprocessed data tables for a strictly controlled comparison. Each baseline is trained from scratch with three independent random seeds (matching the \our{} protocol) and reported as mean$\pm$std.

\subsection{Main comparison}
\label{sec:main}

\begin{table*}[t]
\centering
\caption{\textbf{Main comparison: Pearson R per split} on LP-PDBBind 3-tier hold-out, CASF-2016, and CASF-2016 non-train (161 entries not in LP train). All entries are mean$\pm$std over three independent random seeds (validation RMSE used for model selection within each seed; no hand-picked best run). Best mean per column in \textbf{bold}; second best \underline{underlined}. \our{}-\textsc{Pocket} uses FASTA, pocket-residue list, and SMILES; \our{}-\textsc{NoPocket} removes the pocket annotation; \our{}-\textsc{Pocket-NoESM} keeps the pocket marker but replaces ESM-2 with learned residue embeddings.}
\label{tab:main_r}
\setlength{\tabcolsep}{2pt}
\scriptsize
\begin{tabular}{l c c c c c c}
\toprule
Method & val & test\_cl1 & test\_cl2 & test\_cl3 & CASF-2016 & CASF non-train \\
\midrule
DeepDTA~\cite{deepdta}         & $0.504 \pm 0.013$ & $0.522 \pm 0.009$ & $0.521 \pm 0.017$ & \underline{$0.483 \pm 0.015$} & $0.686 \pm 0.023$ & $0.586 \pm 0.014$ \\
Pafnucy~\cite{pafnucy}         & $0.422 \pm 0.010$ & $0.490 \pm 0.010$ & $0.492 \pm 0.019$ & $0.475 \pm 0.024$ & $0.658 \pm 0.019$ & $0.626 \pm 0.018$ \\
DeepDTAF~\cite{deepdtaf}       & \underline{$0.531 \pm 0.007$} & \underline{$0.531 \pm 0.015$} & \underline{$0.524 \pm 0.026$} & $0.480 \pm 0.037$ & \underline{$0.727 \pm 0.019$} & \underline{$0.638 \pm 0.020$} \\
DEAttentionDTA~\cite{deattentiondta} & $0.516 \pm 0.022$ & $0.516 \pm 0.022$ & $0.507 \pm 0.038$ & $0.456 \pm 0.050$ & $0.697 \pm 0.011$ & $0.618 \pm 0.040$ \\
Cross-modal MS~\cite{dba2025}  & $0.516 \pm 0.008$ & $0.503 \pm 0.011$ & $0.496 \pm 0.010$ & $0.453 \pm 0.014$ & $0.705 \pm 0.009$ & $0.600 \pm 0.046$ \\
\midrule
\textbf{\our{}-\textsc{Pocket}} & $\boldsymbol{0.548 \pm 0.011}$ & $0.507 \pm 0.014$ & $0.496 \pm 0.021$ & $0.433 \pm 0.023$ & $\boldsymbol{0.747 \pm 0.031}$ & $\boldsymbol{0.646 \pm 0.027}$ \\
\textbf{\our{}-\textsc{NoPocket}} & $0.506 \pm 0.003$ & $0.525 \pm 0.027$ & $0.503 \pm 0.031$ & $0.455 \pm 0.033$ & $0.685 \pm 0.043$ & $0.567 \pm 0.017$ \\
\textbf{\our{}-\textsc{Pocket-NoESM}} & $0.529 \pm 0.015$ & $\boldsymbol{0.531 \pm 0.033}$ & $\boldsymbol{0.538 \pm 0.059}$ & $\boldsymbol{0.497 \pm 0.076}$ & $0.713 \pm 0.039$ & $0.632 \pm 0.044$ \\
\bottomrule
\end{tabular}
\end{table*}

\begin{table*}[t]
\centering
\caption{\textbf{RMSE} ($\downarrow$) per split, mean$\pm$std over three seeds. Best per column in \textbf{bold}; second best \underline{underlined}.}
\label{tab:main_rmse}
\setlength{\tabcolsep}{2pt}
\scriptsize
\begin{tabular}{l c c c c c c}
\toprule
Method & val & test\_cl1 & test\_cl2 & test\_cl3 & CASF-2016 & CASF non-train \\
\midrule
DeepDTA~\cite{deepdta}         & $1.432 \pm 0.025$ & $1.476 \pm 0.002$ & $1.597 \pm 0.003$ & $1.626 \pm 0.014$ & $1.597 \pm 0.054$ & $1.727 \pm 0.025$ \\
Pafnucy~\cite{pafnucy}         & $1.489 \pm 0.008$ & $1.512 \pm 0.016$ & $1.636 \pm 0.017$ & $1.615 \pm 0.014$ & $1.801 \pm 0.056$ & $1.764 \pm 0.037$ \\
DeepDTAF~\cite{deepdtaf}       & \underline{$1.390 \pm 0.005$} & $\boldsymbol{1.456 \pm 0.022}$ & \underline{$1.575 \pm 0.041$} & \underline{$1.597 \pm 0.065$} & \underline{$1.525 \pm 0.055$} & \underline{$1.641 \pm 0.027$} \\
DEAttentionDTA~\cite{deattentiondta} & $1.416 \pm 0.014$ & $1.488 \pm 0.046$ & $1.626 \pm 0.084$ & $1.661 \pm 0.120$ & $1.614 \pm 0.057$ & $1.712 \pm 0.025$ \\
Cross-modal MS~\cite{dba2025}  & $1.410 \pm 0.027$ & $1.506 \pm 0.047$ & $1.632 \pm 0.061$ & $1.653 \pm 0.085$ & $1.592 \pm 0.017$ & $1.730 \pm 0.094$ \\
\midrule
\textbf{\our{}-\textsc{Pocket}} & $\boldsymbol{1.381 \pm 0.012}$ & $1.506 \pm 0.025$ & $1.620 \pm 0.038$ & $1.666 \pm 0.060$ & $\boldsymbol{1.492 \pm 0.063}$ & $\boldsymbol{1.618 \pm 0.019}$ \\
\textbf{\our{}-\textsc{NoPocket}} & $1.430 \pm 0.015$ & $1.474 \pm 0.041$ & $1.608 \pm 0.044$ & $1.646 \pm 0.052$ & $1.611 \pm 0.074$ & $1.770 \pm 0.024$ \\
\textbf{\our{}-\textsc{Pocket-NoESM}} & $1.392 \pm 0.013$ & \underline{$1.459 \pm 0.042$} & $\boldsymbol{1.551 \pm 0.070}$ & $\boldsymbol{1.557 \pm 0.069}$ & $1.574 \pm 0.110$ & $1.652 \pm 0.066$ \\
\bottomrule
\end{tabular}
\end{table*}

\begin{table*}[t]
\centering
\caption{\textbf{Concordance Index} ($\uparrow$) per split, mean$\pm$std over three seeds. Best per column in \textbf{bold}; second best \underline{underlined}.}
\label{tab:main_ci}
\setlength{\tabcolsep}{2pt}
\scriptsize
\begin{tabular}{l c c c c c c}
\toprule
Method & val & test\_cl1 & test\_cl2 & test\_cl3 & CASF-2016 & CASF non-train \\
\midrule
DeepDTA~\cite{deepdta}         & $0.670 \pm 0.008$ & $0.671 \pm 0.003$ & $0.674 \pm 0.006$ & $\boldsymbol{0.657 \pm 0.005}$ & $0.743 \pm 0.013$ & $0.701 \pm 0.010$ \\
Pafnucy~\cite{pafnucy}         & $0.641 \pm 0.001$ & $0.660 \pm 0.003$ & $0.656 \pm 0.009$ & $0.650 \pm 0.010$ & $0.733 \pm 0.008$ & $0.713 \pm 0.005$ \\
DeepDTAF~\cite{deepdtaf}       & \underline{$0.683 \pm 0.000$} & $\boldsymbol{0.675 \pm 0.006}$ & \underline{$0.675 \pm 0.011$} & $0.656 \pm 0.016$ & \underline{$0.763 \pm 0.011$} & \underline{$0.722 \pm 0.011$} \\
DEAttentionDTA~\cite{deattentiondta} & $0.677 \pm 0.005$ & $0.669 \pm 0.009$ & $0.667 \pm 0.015$ & $0.645 \pm 0.018$ & $0.750 \pm 0.007$ & $0.708 \pm 0.019$ \\
Cross-modal MS~\cite{dba2025}  & $0.671 \pm 0.004$ & $0.659 \pm 0.005$ & $0.657 \pm 0.002$ & $0.640 \pm 0.004$ & $0.749 \pm 0.007$ & $0.701 \pm 0.018$ \\
\midrule
\textbf{\our{}-\textsc{Pocket}} & $\boldsymbol{0.686 \pm 0.005}$ & $0.665 \pm 0.003$ & $0.659 \pm 0.006$ & $0.633 \pm 0.006$ & $\boldsymbol{0.773 \pm 0.017}$ & $\boldsymbol{0.727 \pm 0.013}$ \\
\textbf{\our{}-\textsc{NoPocket}} & $0.672 \pm 0.004$ & \underline{$0.672 \pm 0.010$} & $0.662 \pm 0.011$ & $0.639 \pm 0.013$ & $0.744 \pm 0.019$ & $0.696 \pm 0.012$ \\
\textbf{\our{}-\textsc{Pocket-NoESM}} & $0.681 \pm 0.006$ & $0.672 \pm 0.010$ & $\boldsymbol{0.675 \pm 0.021}$ & \underline{$0.657 \pm 0.026$} & $0.758 \pm 0.016$ & $0.722 \pm 0.017$ \\
\bottomrule
\end{tabular}
\end{table*}

Tables~\ref{tab:main_r}--\ref{tab:main_ci} report Pearson R, RMSE, and concordance index (CI) for the five reproduced baselines and the three \our{} variants. The results are not dominated by a single configuration. On the validation and CASF-style splits, \our{}-\textsc{Pocket} has the highest mean Pearson R and CI, improving over the closest non-\our{} competitor, DeepDTAF, by $0.017$ on validation, $0.020$ on CASF-2016, and $0.008$ on the CASF-2016 non-train subset. The last margin is small relative to seed variability and should be interpreted as a best-mean observation rather than a substantial improvement.

The strict LP-PDBBind tiers show the opposite pattern. \our{}-\textsc{Pocket-NoESM}, which keeps the pocket marker but replaces ESM-2 with a learned 21-vocabulary residue embedding, obtains the highest mean Pearson R on all three no-leak tiers: it ties DeepDTAF on \texttt{test\_cl1} and exceeds it by $0.014$ on both \texttt{test\_cl2} and \texttt{test\_cl3}. It also gives the lowest RMSE on \texttt{test\_cl2}/\texttt{test\_cl3}. Meanwhile, \our{}-\textsc{NoPocket} consistently improves over \our{}-\textsc{Pocket} on the strict LP tiers. Together, these trends indicate that both added priors--the pocket marker and the frozen PLM representation--can become counterproductive under no-leak evaluation, so the appropriate \our{} variant depends on the deployment regime rather than on a global ranking.

Because several margins are small, we treat best-mean rankings as descriptive unless supported by paired resampling. We average predictions across the three seeds for each method and compute paired bootstrap intervals. These intervals show that \our{}-\textsc{Pocket-NoESM} improves over \our{}-\textsc{Pocket} on all three strict LP tiers (bootstrap $\Delta$R $=+0.037,+0.060,+0.082$, all intervals $>0$ and all RMSE deltas $<0$), and improves over DeepDTAF on \texttt{test\_cl2}/\texttt{test\_cl3} but not on \texttt{test\_cl1}. The familiar-split gains of \our{}-\textsc{Pocket} over DeepDTAF remain descriptive because the 95\% intervals cross zero on validation, CASF-2016, and CASF-2016 non-train.

\begin{figure*}[t]
  \centering
  \includegraphics[width=0.95\linewidth]{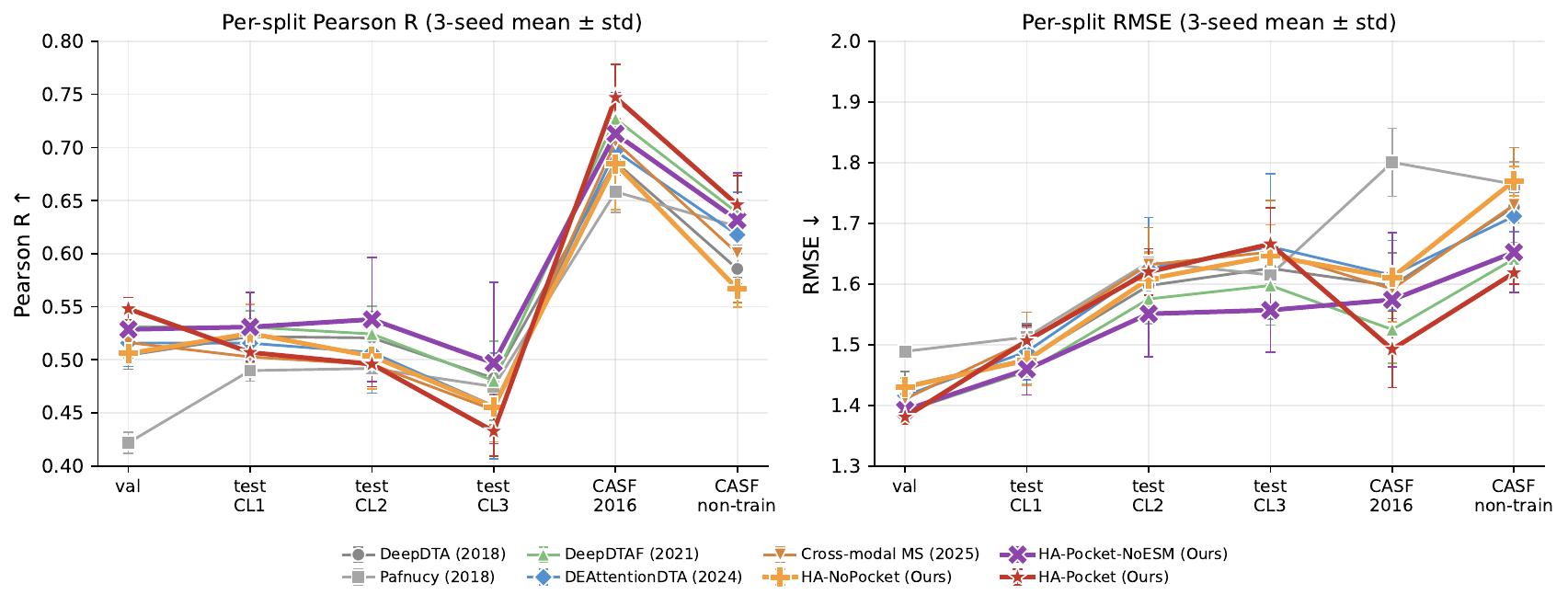}
  \caption{\textbf{Per-split generalization curves} (3-seed mean $\pm$ std error bars). Pearson R (left, higher better) and RMSE (right, lower better) are shown across the six evaluation splits. \our{}-\textsc{Pocket} (red) leads on validation and the two CASF splits; \our{}-\textsc{Pocket-NoESM} (purple) leads on the strict LP no-leak tiers (\texttt{cl1}--\texttt{cl3}); \our{}-\textsc{NoPocket} (orange) is the variant for targets without pocket annotations. The strict tiers show the paired reversal analyzed in Sections~\ref{sec:ablation_esm}--\ref{sec:ablation_pock}: removing either ESM-2 or the pocket marker improves the corresponding no-leak comparison.}
  \label{fig:gen}
\end{figure*}

\subsection{Ablation: frozen ESM-2 vs. learned residue embeddings}
\label{sec:ablation_esm}

\begin{table}[t]
\centering
\caption{\textbf{ESM-2 input vs.\ learned residue embedding} on the \our{}-\textsc{Pocket} architecture: this comparison serves both as an ablation and as the third deployable variant. Same schedule, same data, same pocket marker; only the protein token representation is swapped (ESM-2 $\to$ 21-vocabulary residue embedding). Pearson R is reported as mean$\pm$std over three seeds. $\Delta\mathrm{R} =$ (\our{}-\textsc{Pocket-NoESM}) $-$ (\our{}-\textsc{Pocket}); positive values indicate that replacing ESM-2 improves R on that split. ESM-2 helps in-distribution (val, CASF-2016, CASF-2016 non-train) but reduces R on every strict LP no-leak tier, producing a second split-conditioned reversal parallel to the pocket-marker reversal in Table~\ref{tab:ablation_pock}. The right column (\our{}-\textsc{Pocket-NoESM}) is therefore the preferred \our{} configuration when pocket annotations are available under strict LP-style deployment.}
\label{tab:ablation_esm}
\setlength{\tabcolsep}{3pt}
\scriptsize
\begin{tabular}{l c c c c}
\toprule
Split & \textsc{Pocket} & \textsc{Pocket-NoESM} & $\Delta$R & rel. \\
\midrule
val                & $\boldsymbol{0.548 \pm 0.011}$ & $0.529 \pm 0.015$ & $-0.019$ & $-3.5\%$ \\
test\_cl1          & $0.507 \pm 0.014$ & $\boldsymbol{0.531 \pm 0.033}$ & $+0.024$ & $+4.7\%$ \\
test\_cl2          & $0.496 \pm 0.021$ & $\boldsymbol{0.538 \pm 0.059}$ & $+0.042$ & $+8.5\%$ \\
test\_cl3          & $0.433 \pm 0.023$ & $\boldsymbol{0.497 \pm 0.076}$ & $+0.064$ & $+14.9\%$ \\
\textbf{CASF-2016}          & $\boldsymbol{0.747 \pm 0.031}$ & $0.713 \pm 0.039$ & $-0.034$ & $-4.6\%$ \\
CASF non-train     & $\boldsymbol{0.646 \pm 0.027}$ & $0.632 \pm 0.044$ & $-0.014$ & $-2.2\%$ \\
\bottomrule
\end{tabular}
\end{table}

To isolate the contribution of pretrained protein context, we re-train the \our{}-\textsc{Pocket} architecture with a 21-vocabulary residue embedding replacing the ESM-2 features, a configuration we hereafter call \our{}-\textsc{Pocket-NoESM} and treat as a third deployable variant rather than only as a diagnostic ablation (Table~\ref{tab:ablation_esm}). The data reveal a second split-conditioned reversal that mirrors the pocket-marker reversal of Section~\ref{sec:ablation_pock}: ESM-2 improves Pearson R on the canonical / familiar splits (val: $\Delta=+0.019$ in favor of ESM, CASF-2016: $+0.034$, CASF-2016 non-train: $+0.014$), but reduces Pearson R on every strict LP no-leak tier ($-0.024$ on \texttt{test\_cl1}, $-0.042$ on \texttt{test\_cl2}, $-0.064$ on \texttt{test\_cl3}). Cross-referencing Table~\ref{tab:main_r}, \our{}-\textsc{Pocket-NoESM} is the highest-mean-R \our{} variant on every strict LP tier, matching DeepDTAF on \texttt{cl1} and exceeding it on \texttt{cl2}/\texttt{cl3}. The mechanism we propose is consistent with the Discussion (Section~\ref{sec:disc}): PLM features primarily encode within-family discriminators, including residue conservation, secondary-structure cues, and binding-site signatures, that are highly informative when the test target's family overlaps the training distribution but can become uninformative or misleading when LP similarity filtering removes that overlap. The fact that two structurally unrelated components exhibit the same canonical-vs-LP reversal supports our view that this pattern is tied to the evaluation regime rather than to either component alone.

\subsection{Ablation: pocket-position marker}
\label{sec:ablation_pock}

\begin{table}[t]
\centering
\caption{\textbf{\our{}-\textsc{Pocket} vs.\ \our{}-\textsc{NoPocket}: the split-conditioned reversal.} Pearson R, mean$\pm$std over three seeds. The pocket-position marker improves R on the canonical / familiar splits (val, CASF-2016, CASF-2016 non-train) but reduces R on every strict LP no-leak tier. We therefore keep the no-pocket variant alongside the pocket-equipped variants and choose among configurations by deployment regime rather than collapsing them into a single default.}
\label{tab:ablation_pock}
\setlength{\tabcolsep}{2pt}
\scriptsize
\begin{tabular}{l c c c}
\toprule
Split & \textsc{Pocket} & \textsc{NoPocket} & $\Delta$R \\
\midrule
val                & $\boldsymbol{0.548 \pm 0.011}$ & $0.506 \pm 0.003$ & $-0.042$ \\
test\_cl1          & $0.507 \pm 0.014$ & $\boldsymbol{0.525 \pm 0.027}$ & $+0.018$ \\
test\_cl2          & $0.496 \pm 0.021$ & $\boldsymbol{0.503 \pm 0.031}$ & $+0.008$ \\
test\_cl3          & $0.433 \pm 0.023$ & $\boldsymbol{0.455 \pm 0.033}$ & $+0.023$ \\
\textbf{CASF-2016}          & $\boldsymbol{0.747 \pm 0.031}$ & $0.685 \pm 0.043$ & $-0.063$ \\
CASF non-train     & $\boldsymbol{0.646 \pm 0.027}$ & $0.567 \pm 0.017$ & $-0.079$ \\
\bottomrule
\end{tabular}
\end{table}

Table~\ref{tab:ablation_pock} compares the ESM-equipped variants with and without the pocket marker. The marker improves R on the canonical splits where the test target's protein and pocket geometry are likely represented in training (val, CASF-2016, CASF-2016 non-train), and reduces R on all three strict LP no-leak tiers, where the test pocket is by construction dissimilar from anything in training. We interpret this as the marker providing an inductive bias toward training-distribution pocket geometry; that bias is informative when the test pocket is familiar and can be misleading when it is not. Combined with the parallel ESM reversal of Section~\ref{sec:ablation_esm}, this yields the three deployment-regime-matched recommendations summarized in the introduction: \our{}-\textsc{Pocket} for familiar targets with a pocket annotation, \our{}-\textsc{Pocket-NoESM} for strict LP-style targets with a pocket annotation, and \our{}-\textsc{NoPocket} for targets without a pocket annotation. The reversal itself would have been concealed under canonical-only or LP-only reporting, which is why we treat paired split evaluation as a methodological requirement rather than a supplementary check.

\subsection{Variant dynamics and CASF-2016 scatter}

\begin{figure}[t]
  \centering
  \includegraphics[width=\linewidth]{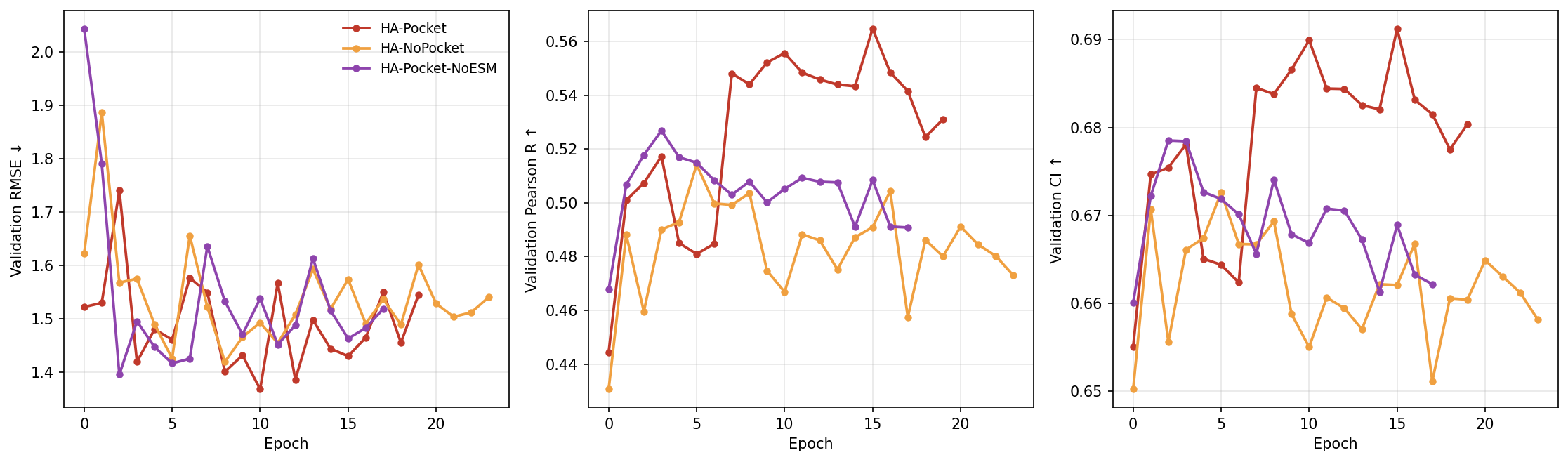}
  \caption{\textbf{Training dynamics of the three \our{} variants.} Validation RMSE / Pearson R / CI per epoch for \our{}-\textsc{Pocket}, \our{}-\textsc{NoPocket}, and \our{}-\textsc{Pocket-NoESM}, shown for single representative seeds. This figure is diagnostic rather than a model-selection result: the final comparison uses validation-RMSE-selected checkpoints and three-seed mean$\pm$std reporting in Tables~\ref{tab:main_r}--\ref{tab:main_ci}.}
  \label{fig:curves}
\end{figure}

\begin{figure}[t]
  \centering
  \includegraphics[width=\linewidth]{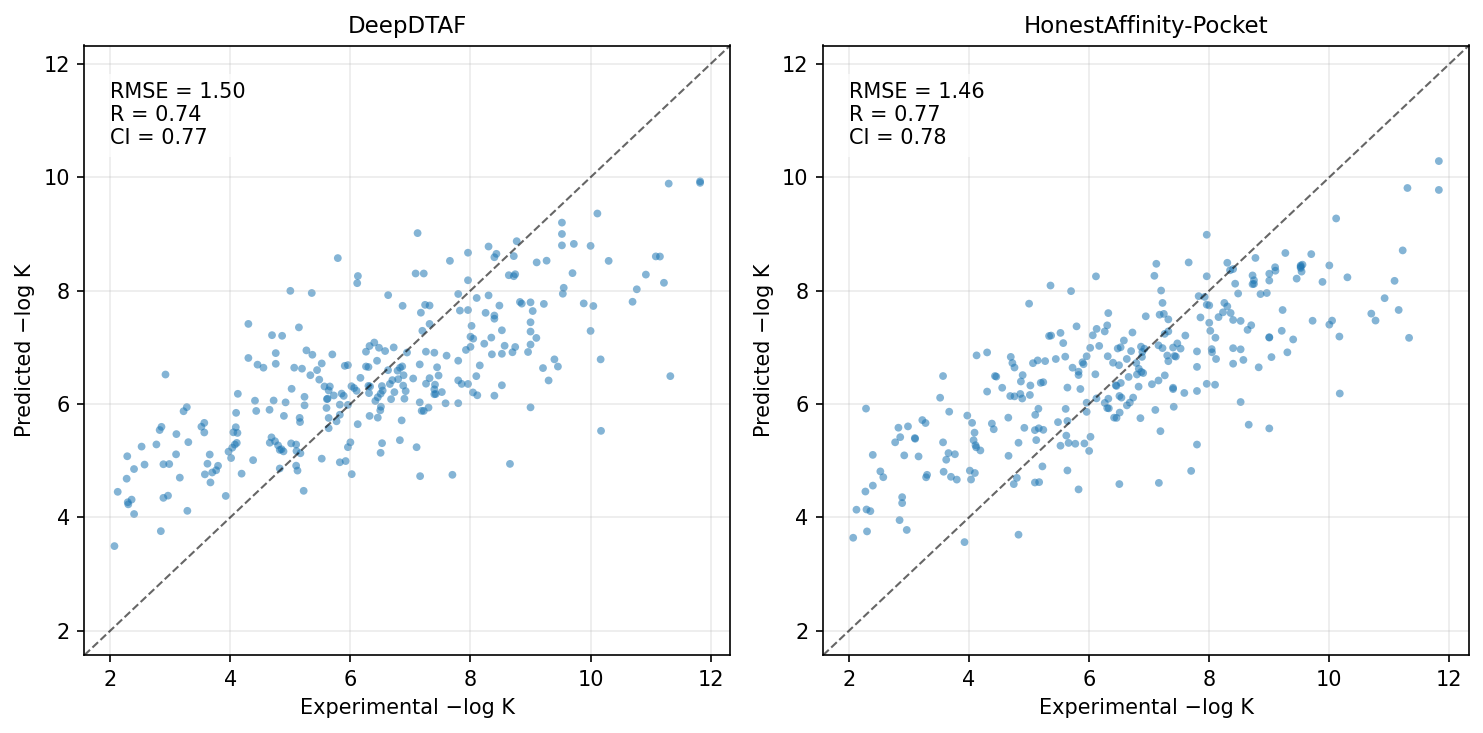}
  \caption{\textbf{CASF-2016 scatter.} Seed-averaged predicted vs.\ experimental $-\log K$ for DeepDTAF~\cite{deepdtaf} (left), the strongest non-\our{} CASF competitor in Table~\ref{tab:main_r}, and \our{}-\textsc{Pocket} (right). The scatter is qualitative; the numerical comparison and paired uncertainty check are reported in Tables~\ref{tab:main_r}--\ref{tab:main_ci} and Section~\ref{sec:main}.}
  \label{fig:scatter}
\end{figure}

Fig.~\ref{fig:curves} visualizes the validation trajectories of the three deployable \our{} variants, supporting the view that the variants are not separate model families but controlled switches over the same template. Fig.~\ref{fig:scatter} provides a qualitative CASF-2016 view against DeepDTAF, the strongest non-\our{} CASF competitor in Table~\ref{tab:main_r}; the formal claims remain the three-seed tables and paired bootstrap analysis rather than a single scatter panel.

\section{Discussion}
\label{sec:disc}

The frozen ESM-2 input improves Pearson R on the canonical / familiar splits (val, CASF-2016, CASF-2016 non-train), but the $-$ESM ablation in Section~\ref{sec:ablation_esm} shows that the same PLM input reduces Pearson R by $0.024$--$0.064$ on every strict LP no-leak tier. This pattern is consistent with PLM features encoding within-family discriminators, such as residue conservation, secondary-structure cues, and binding-site signatures, that are most informative when the test target's protein family overlaps the training distribution. Under LP-style similarity filtering, that overlap is reduced by construction, and signals that were useful within-family discriminators can become correlated with target identity rather than ligand affinity. The corollary is that out-of-distribution generalization to unfamiliar families will likely require ligand-side foundation models, explicit 3D inductive biases, or both.

A central methodological finding is that two structurally unrelated architectural choices, the binary pocket-position marker and the 1280-dimensional frozen ESM-2 input, both change sign between canonical and leak-proof evaluations in the same direction: they improve in-distribution performance and reduce strict LP performance. Single-regime reporting would have concealed both reversals. Canonical-only evaluation would favor the pocket marker and ESM input without revealing their strict-tier cost, whereas strict-LP-only evaluation would discount components that are useful for in-distribution scoring on familiar targets. The mechanism is similar in both cases: each component provides an inductive bias toward training-distribution structure, either pocket geometry or protein-family signatures. That bias is informative when the test target is familiar and can be misleading when it is not. We therefore recommend paired canonical / leak-proof ablations as a routine sanity check for new affinity scorers, and we keep all three variants of \our{} (\textsc{Pocket}, \textsc{NoPocket}, \textsc{Pocket-NoESM}) rather than picking a single best configuration.

Relative to structure-aware scorers, \our{} occupies a different deployment regime. Methods that consume 3D complex geometry, including voxel CNNs, atomic GNNs, and SE(3)-equivariant networks, typically reach Pearson R $> 0.85$ on canonical CASF-2016 when high-quality crystal poses are available. \our{}'s 1D-input numbers do not close that gap, as expected. The trade-off is deployment scope: \our{}-\textsc{NoPocket} requires only the protein FASTA and a SMILES string at inference, while \our{}-\textsc{Pocket} and \our{}-\textsc{Pocket-NoESM} additionally require a deposited pocket-residue list, the same input regime as DeepDTAF, but no docking or structure-prediction pipeline. All three variants train on 11{,}513 leak-proof complexes in $\sim$3 GPU-hours on a single Tesla V100 and infer in $\sim$10\,ms per complex once the embedding cache is warm. We view 1D-input PLM-augmented scoring and structure-aware scoring as complementary rather than competing regimes.

Several limitations remain. \our{}-\textsc{Pocket} and \our{}-\textsc{Pocket-NoESM} both require a deposited pocket residue list; for screening targets without one, \our{}-\textsc{NoPocket} is the appropriate variant. On the strict LP no-leak tiers, \our{}-\textsc{NoPocket} is stronger than \our{}-\textsc{Pocket}, but \our{}-\textsc{Pocket-NoESM} is preferred when pocket annotations are available. ESM-2 weights are frozen throughout, and parameter-efficient adaptation such as LoRA may recover part of the strict-tier gap at higher training cost. The SMILES branch is also trained from scratch; replacing it with a ligand foundation model such as Uni-Mol~\cite{unimol} or MolFormer~\cite{molformer} is a natural extension. We omit CASF-2013 because the canonical PDB list was not bundled in the v2020R1 redistribution we used. Finally, although we report mean$\pm$std across three random seeds and provide paired bootstrap checks over complexes for selected comparisons, small best-mean margins, especially on CASF-2016 non-train, should be read descriptively.

\section{Conclusion}
\label{sec:conclusion}

We presented \our{}, a 1D-input protein--ligand affinity predictor that injects frozen ESM-2 (650M) embeddings and an optional pocket-position marker into a compact multi-scale convolution / matrix-product scoring template, trained on 11{,}513 leak-proof complexes in $\sim$3 GPU-hours on a single Tesla V100 and reported as mean$\pm$std over three random seeds. Rather than collapse the architecture to a single best configuration, we report three variants of the same template, each matched to a distinct deployment regime: \our{}-\textsc{Pocket} for familiar / annotated targets, \our{}-\textsc{Pocket-NoESM} for strict LP-style targets with pocket annotations, and \our{}-\textsc{NoPocket} for targets without pocket annotations. The paper's primary methodological message is that architectural components can change sign between canonical and leak-proof evaluations: in our experiments, both the pocket-position marker and the 1280-dimensional frozen ESM-2 input help in-distribution and reduce Pearson R on every strict LP no-leak tier. Single-regime reporting would have concealed both reversals. These results support leak-proof splits as a default reporting regime, paired canonical / leak-proof ablations as a standard diagnostic for new affinity scorers, and multi-variant reporting when deployment regimes differ. We provide an anonymous code snapshot and preprocessing scripts through the first-page footnote; pretrained checkpoints will be released upon acceptance. PDBbind-derived data artifacts will be reproducible from the original licensed sources rather than redistributed directly.

\bibliographystyle{IEEEtran}
\bibliography{refs}

\end{document}